\documentclass[conference,onecolumn]{IEEEtran}
\ifCLASSINFOpdf
\else
\fi

\usepackage{amsmath, amssymb}

\usepackage{tikz}
\usetikzlibrary{shapes,arrows,fit}
\usepackage{pgfplots}
 \pgfplotsset{compat=newest} 
 \pgfplotsset{plot coordinates/math parser=false}
\usepackage[none]{hyphenat}
\usepackage[normalem]{ulem} 
\usepackage{cite}
\usepackage{graphicx}

\newcommand{\tone}{\ensuremath{T_1}}
\newcommand{\ttwo}{\ensuremath{T_2}}
\newcommand{\tr}{\ensuremath{r}}

\newcommand{\wone}{\ensuremath{W_1}}
\newcommand{\wtwo}{\ensuremath{W_2}}

 \newcommand{\mc}[1]{\mathcal{#1}}

\newcommand{\ve}[1]{\uline{\smash #1}}

\newcommand{\hY}{\hat Y_r}
\newcommand{\hy}{\hat y_r}
\newcommand{\hyri}{\hat y_{ri}}

\newcommand{\yhcard}{\ensuremath{L}}
\newcommand{\ycard}{\ensuremath{|\mc Y_r |}}
\DeclareMathOperator{\dom}{\mathbf{dom}}

\newtheorem{prop}{Proposition}
\newtheorem{lemma}{Lemma}
\newtheorem{cor}{Corollary}

\tikzstyle{block2} = [draw, rectangle, 
    minimum height=1.5em, minimum width=2em]
\tikzstyle{sum} = [draw, circle, inner sep=0pt, node distance=1cm]

\begin{document}

\title{Rate-Distortion Properties of Single-Layer Quantize-and-Forward for Two-Way Relaying}

\author{\IEEEauthorblockN{ Michael Heindlmaier, Onurcan \.{I}\c{s}can}

\IEEEauthorblockA{Institute for Communications Engineering, Technische Universit\"at M\"unchen,
Munich, Germany\\
Email: michael.heindlmaier@tum.de, onurcan.iscan@tum.de}}

\maketitle
\thispagestyle{plain}
\pagestyle{plain}

\begin{abstract}
The Quantize \& Forward (QF) scheme for two-way relaying is studied with a focus on its rate-distortion properties.
A sum rate maximization problem is formulated and the associated quantizer optimization problem is investigated.
An algorithm to approximately solve the problem is proposed. Under certain cases scalar quantizers maximize the sum rate.
\end{abstract}  

\section{Introduction}
Consider a communication system where two user nodes \tone\, and \ttwo\, exchange their messages with the help of a relay and there is no direct link between \tone\, and \ttwo. This scenario is known as a separated two-way relay channel \cite{4797715,rankov2006achievable}. 
In this work, we focus on a specific version of Quantize \& Forward (QF) relaying \cite{schnurr2007achievable}: the relay maps its received (noisy) signal to a quantization index by using a quantizer function $\mathcal{Q}(.)$. The index is then digitally transmitted to the destination nodes through the downlink channels. Compared to our previous work \cite{6620441} where we studied QF relaying schemes under symmetric conditions, this scheme is more flexible for asymmetric setups and exploits the correlation of the quantization index with the users' symbols.
Our main contribution is the derivation of some properties of the corresponding rate-distortion problem. We further propose an algorithm to obtain quantizer distributions that maximize the sum rate and serve as a starting point for quantizer design.
\section{System Model}
\label{sec:sysmod}
\begin{figure}
\centering
\includegraphics[width=0.45\textwidth]{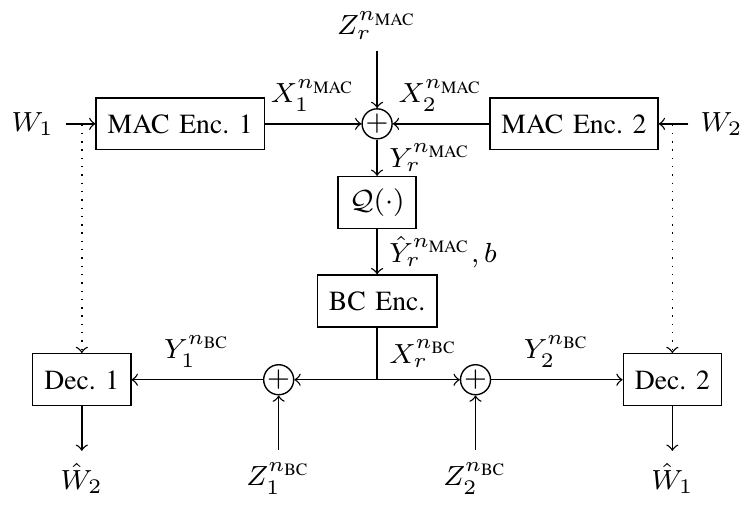}
\caption{System Model}
\label{fig:block}
\end{figure}

Two source nodes \tone\, and \ttwo\, exchange their messages $\wone\ \in \{1,2,\ldots, 2^{nR_1}\}$, $\wtwo\ \in \{1,2,\ldots, 2^{nR_2}\}$ in $n$ channel uses through a relay node \tr. \tone\, and \ttwo\, cannot hear each other and have to communicate over the relay. The communication consists of two phases: In the multiple access (MAC) phase with $n_\text{MAC}$ channel uses, \tone\, and \ttwo\, encode their messages \wone\, and \wtwo\ to the channel inputs $X_1^{n_\text{MAC}}$  and $X_2^{n_\text{MAC}}$, respectively, with $X_{1,t} \in \mc X_1$, $X_{2,t} \in \mc X_2$. 
 The time fraction of this first phase is $\alpha = n_\text{MAC} / n$.
The relay receives
 \begin{equation}
  Y_{r,t} = X_{1,t} + X_{2,t} + Z_{r,t}, \quad t=\{1,2,\ldots,n_\text{MAC}\},
 \end{equation}
where $Z_{r,t} \sim \mc N(0,N_r)$ and $\mathbb E\{X_{1,t}^2 \} \leq P_1$, $\mathbb E\{X_{2,t}^2 \} \leq P_2$.
The relay maps $Y_r^{n_\text{MAC}}$ to a single quantized representation $\hY^{n_\text{MAC}}$ with symbol alphabet $\mc{\hat Y}_r$. Denote the number of possible quantizer levels as $L=|\mc{\hat Y}_r|$ and the associated quantizer index as $b = \mc Q(Y_r^{n_\text{MAC}})$. 
During the Broadcast (BC) phase with $n_\text{BC} = n-n_\text{MAC}$ channel uses, the relay transmits the codeword $X_r^{n_\text{BC}}(b)$.
The received signals at \tone\, and \ttwo\, are
\begin{align}
Y_{j,t} = X_{r,t} + Z_{j,t} , \quad t=\{n_\text{MAC}+1,\ldots,n\},
\end{align}
for $j\in\{1,2\}$, $\mathbb E\{X_{r,t}^2 \} \leq P_r$ and $Z_{j,t} \sim \mc N(0,N_j)$.
Nodes \tone\, and \ttwo\ decode $W_2$ and $W_1$, respectively, by using their own message as side information. Fig. \ref{fig:block} depicts the system setup.
In the following, we omit the time index $t$ if we refer to a single channel use.

\section{Rate-Distortion Properties}
\subsection{Achievable Rates}
\begin{figure}
\centering
\includegraphics[width=0.25\textwidth]{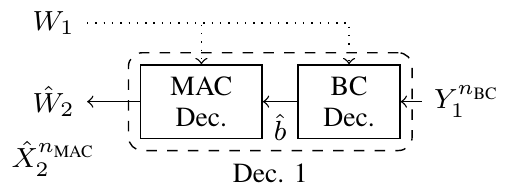}
\caption{Decoder structure}
\label{fig:decCF}
\end{figure}

Different coding schemes have been proposed for the setup of Fig.~\ref{fig:block}. A summary can be found in \cite{6620441}. In this work we focus on the approach of \cite{schnurr2007achievable} that requires reliable decoding of the quantization index $b$ at both receivers. One can exploit the fact that $\hY$ is correlated with $X_1$ and $X_2$. Similar to the approach in \cite{tuncel2006slepian}, one reliably transmits the quantization index $b$ to both users, trading off correlation for channel quality. The BC code is decoded using the own message as a priori knowledge. Knowing $b$, the desired message is decoded, again using the own message as side information.
The structure of this decoder is shown in Fig. \ref{fig:decCF}.
The achievable rate region \cite{schnurr2007achievable,kim2008comparison} is the set $\mc R~\subset~ \mathbb R_+^2$ of rate tuples $(R_1,R_2)$ satisfying
\begin{align}
&R_1 \leq \alpha I(X_1;\hY|X_2,U), \qquad R_2 \leq \alpha I(X_2;\hY|X_1,U) \nonumber 
\\
&\alpha I(Y_r ; \hY | X_2,U) \leq (1-\alpha) I(X_r; Y_2) = (1-\alpha)I_2 \nonumber \\
&\alpha I(Y_r ; \hY | X_1,U) \leq (1-\alpha) I(X_r; Y_1) = (1-\alpha)I_1 
\end{align}
for some $p(u)p(x_1|u)p(x_2|u)p(y_r|x_1, x_2)p(\hy|y_r)$ and $p(x_r)p(y_1,y_2|x_r)$, $\alpha > 0$. 
It suffices to consider $|\mc U|\leq 4$, $|\mc{\hat Y}_r| \leq |\mc Y_r| + 3$.

\subsection{Sum Rate Optimization}
\begin{figure}[t]
\centering
\includegraphics[width=0.25\textwidth]{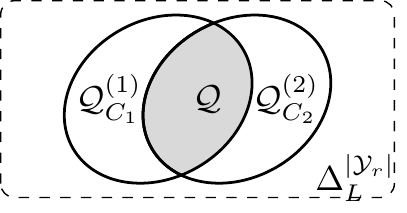}
\caption{Illustration of the feasible set $\mc Q$, a subset of $\Delta_{\yhcard}^{\ycard}$.}
\label{fig:setsQCs}
\end{figure}

$\mc R$ depends on the conditional probability mass function (pmf) $p(\hy|y_r)$ that is induced by the quantizer. We seek for $p(\hy|y_r)$ and an optimal time sharing coefficient $\alpha$ that maximize the sum rate $R_1 + R_2$. By setting $\mc U = \emptyset$, we have the following problem with optimization variables $\alpha$ and $p(\hy|y_r)$:
\begin{alignat}{2}
  &\text{maximize  } \qquad &\alpha \cdot \left( I(X_1;\hY|X_2) + I(X_2;\hY|X_1) \right)  \nonumber \\
  &\text{subject to:}   &\alpha I(Y_r ; \hY | X_2) \leq  (1-\alpha) I_2 \nonumber \\
  & & \alpha I(Y_r ; \hY | X_1) \leq (1-\alpha)I_1 \nonumber \\
  & & 0 <  \alpha < 1 \label{eq:maximize}
\end{alignat}
The distribution $p(\hy|y_r)$ is represented by the stochastic matrix $Q$, where $q_{ij} \triangleq p(\hat{y}_{ri}|y_{rj})$. Both notations will be used interchangeably.
Each column of $Q$ is a distribution on $\mc{\hat Y}_r$, which is in the simplex $\Delta_{L}$ of all $L$-dimensional probability vectors. $Q$ is thus an element of $\Delta_L^{|\mc Y_r|}$.
$I(Y_r;\hY|X_1)$ and $I(Y_r;\hY|X_2)$ are convex in $p(\hy|y_r)$ for a fixed $p(y_r)$ \cite[Theorem 2.7.4]{cover2006elements}, so the sublevel sets 
\begin{align}
 \mc Q_{C_1}^{(1)}:= \left\{Q \in \Delta_L^{|\mc Y_r|} \big| I(Y_r;\hY|X_1) \leq  C_1\right\} \nonumber\\
\mc Q_{C_2}^{(2)}:= \left\{Q \in \Delta_L^{|\mc Y_r|} \big| I(Y_r;\hY|X_2) \leq  C_2 \right\}
 \end{align}
 are convex \cite[Chapter 3.1.6]{boyd2004convex}. 
Convexity is preserved under intersection, so the feasible set $\mc Q = \mc Q_{C_1}^{(1)} \cap \mc Q_{C_2}^{(2)}$ is convex, as illustrated in Fig.~\ref{fig:setsQCs}.
Define the function 
\begin{align}
 I_{\text{RD}}(C_1,C_2) &= \sup_{Q \in \mc Q_{C_1}^{(1)} \cap \mc Q_{C_2}^{(2)} } \left( I(X_1;\hY|X_2) + I(X_2; \hY|X_1) \right). %\nonumber\\
 %\text{s.t. }  & I(Y_r; \hY| X_1) \leq C_1, \quad I(Y_r; \hY| X_2) \leq C_1
\label{eq:def_I(C1C2)}
\end{align}
$I_{\text{RD}}$ characterizes the tradeoff between the quantization rates supported by the downlink and the sum rate\footnote{The corresponding distortion to be minimized is equal to $H(X_1|\hY, X_2) + H(X_2|\hY,X_1)$}.
Note that (\ref{eq:def_I(C1C2)}) is closely related to the sum rate optimization problem in (\ref{eq:maximize}): If the function $I_{\text{RD}}(C_1,C_2)$ is known, then 
\begin{align}
 \left(R_1 + R_2\right)^* = \sup_{\alpha \in (0,1)} \left[\alpha \cdot I_{\text{RD}}\left( \frac{1-\alpha}{\alpha}I_1, \frac{1-\alpha}{\alpha} I_2 \right) \right]
\label{eq:IC1C2_dependent_on_alpha}
\end{align}
is the solution to problem (\ref{eq:maximize}).
\subsection{Properties of $I_{\text{RD}}(C_1,C_2)$}
\paragraph{Upper Bound}
$I_{\text{RD}}(C_1,C_2)$ is upper bounded by  
\begin{align}
I_{\text{RD}}(C_1,C_2) \leq I(X_1;Y_r|X_2) + I(X_2;Y_r|X_1), 
\end{align}
with equality if $C_1 \geq H(Y_r|X_1)$, $C_2 \geq H(Y_r|X_2)$.
\paragraph{$I_{\text{RD}}(C_1,C_2)$ is nondecreasing in $C_1,C_2$}
As $I(X_1;\hY|X_2) + I(X_2; \hY|X_1)$ is convex in $p(\hy|y_r)$, the problem is a maximization of a convex function over a convex set. According to the maximum principle \cite[Cor. 32.3.2]{rockafellar1997convex}, the optimum of the problem is found at the boundary of $\mc Q$. That is, at least one inequality constraint is satisfied with equality.

For $H(Y_r|X_1) \geq C_1' > C_1$ and $H(Y_r|X_2) \geq C_2' > C_2$ it follows that $I_{\text{RD}}(C_1',C_2') > I_{\text{RD}}(C_1,C_2)$.
However, we do not know if $I_{\text{RD}}(C_1,C_2') > I_{\text{RD}}(C_1,C_2)$ or $I_{\text{RD}}(C_1',C_2) > I_{\text{RD}}(C_1,C_2)$ as the feasible set $\mc Q$ might not change even if $\mc Q_{C_1'}^{(1)} \supset \mc Q_{C_1}^{(1)}$ and $\mc Q_{C_2'}^{(2)} \supset \mc Q_{C_2}^{(2)}$.

\paragraph{Concavity 1} 
\label{property:IC1C2concave} 
 $I_{\text{RD}}(C_1,C_2)$ is a concave function in $C_1$ and $C_2$, for $C_1 \leq H(Y_r|X_1)$, $C_2 \leq H(Y_r|X_2)$, and it is sufficient to choose $|\mc{\hat Y}_r| \leq |\mc{Y}_r|+2$.
This is one of our main results. The proof can be found in 
%\cite{heindlmaier2014ext}.
the Appendix.

\paragraph{Concavity 2} If $I_{\text{RD}}$ is twice differentiable, the function $\alpha I_{\text{RD}}\left( \frac{1-\alpha}{\alpha}I_1, \frac{1-\alpha}{\alpha} I_2 \right)$ is concave in $\alpha$ where $C_1$ and $C_2$ are positive constants independent of $\alpha$. The proof is an adaption of \cite[Proposition 1]{6620441} to the 2-dimensional case.

\subsection{Evaluating the Function $I_{\text{RD}}(C_1,C_2)$}
To evaluate the function $I_{\text{RD}}(C_1,C_2)$ in (\ref{eq:def_I(C1C2)}), we write the Lagrangian:
\begin{align}
 \mc L(Q, \lambda_1, \lambda_2, \nu_1, \nu_2, \ldots,\nu_{|\mc Y_r|})= 
  J(Q) - \lambda_1 I(Y_r; \hY|X_1) - \lambda_2 I(Y_r;\hY|X_2) + \sum_{i,j} \nu_j  q_{ij} 
\end{align}
with $J(Q)=I(X_1;\hY|X_2) + I(X_2;\hY|X_1)$
and $\lambda_1,\lambda_2 \geq 0$.
The last term in the Lagrangian is due to the fact that $Q$ is a conditional distribution, so 
\[\sum_{\hy} p(\hy|y_r) = 1 \quad \forall y_r \qquad \Leftrightarrow \qquad \sum_{i} q_{ij} = 1 \quad \forall j.\]
From the KKT conditions \cite{bertsekas2007nonlinear}, we obtain the optimality conditions $\forall$ $i=1,\ldots,|\mc{\hat{Y}}_r|$, $j=1,\ldots,|\mc Y_r|$
\begin{align}
0 \stackrel{!}{=}   \frac{\partial J}{\partial q_{ij}} 
- \lambda_1 \sum_{x_1} p(x_1) p(y_{rj}|x_1) \log \frac{p(\hat y_{ri}|y_{rj})}{p(\hat y_{ri} | x_1)} 
- \lambda_2 \sum_{x_2} p(x_2) p(y_{rj}|x_2) \log \frac{p(\hat y_{ri}|y_{rj})}{p(\hat y_{ri} | x_2)} + \nu_j.
\label{eq:deriv=0}
\end{align}
Similarly to \cite[Section 10]{cover2006elements}, define
\begin{align}
 \log \mu_j :=  \frac{-\nu_j}{(\lambda_1 + \lambda_2) p(y_{rj})}.  
\end{align}
It follows that
\begin{align}
 q_{ij} = \frac{\exp\left(\delta(\hyri,y_{rj})  \right) }{\mu_j} 
\end{align}
with $\delta(\hyri,y_{rj})$ defined as
\begin{align}
 \delta(\hyri,y_{rj}):=\frac{1}{(\lambda_1 + \lambda_2) p(y_{rj}) } \left( \frac{\partial J}{\partial q_{ij}}  + \lambda_1 \sum_{x_1} p(x_1,y_{rj}) \log p(\hyri|x_1) + \lambda_2 \sum_{x_2} p(x_2,y_{rj}) \log p(\hyri|x_2) \right). \label{eq:delta}
\end{align}

Note that $\delta(\hyri,y_{rj})$ is a function of the distributions $p(x_1|\hy,x_2)$, $p(x_2|\hy,x_1)$, $p(\hy|x_1)$ and $p(\hy|x_2)$.
Since $\sum_l q_{lj} = 1$ $\,\forall j$,\, the optimality conditions are
\begin{align}
  q_{ij} = \frac{\exp\left( \delta(\hyri,y_{rj}) \right)  }{\sum_{l}  \exp\left( \delta(\hat{y}_{rl},y_{rj}) \right) } .
  \label{eq:optimality_cond_qij_schnurr}
\end{align}
Eq.~\eqref{eq:optimality_cond_qij_schnurr} is not an explicit characterization of $Q$ because the RHS depends on $Q$ as well.
One can solve for a conditional pmf $Q$ satisfying (\ref{eq:optimality_cond_qij_schnurr}) with the following iterative algorithm:

\begin{enumerate}
\item Choose two Lagrangian multipliers $\lambda_1>0$, $\lambda_2>0$, $\epsilon >0$ and set $k=0$.  
\item Choose an initial conditional pmf $Q^{(0)} \triangleq p^{(0)}(\hy|y_r)$ and calculate $p^{(0)}(x_1|\hy, x_2)$, $p^{(0)}(x_2|\hy, x_1)$, $p^{(0)}(\hy|x_1)$ and $p^{(0)}(\hy|x_2)$ according to $p^{(0)}(\hy|y_r)$.
\item Calculate the value of the Lagrangian\footnote{Note that the conditions for $Q$ being a conditional pmf are always satisfied and the last term in the Lagrangian can be omitted.} $\mc L^{(0)}$ with the current distributions.
\item Increase $k$ by $1$.
\item Calculate $\delta^{(k)}(\hyri,y_{rj})$ $\forall$ $i=1,\ldots,|\mc{\hat{Y}}_r|$, $j=1,\ldots,|\mc Y_r|$, with the current distributions.
\item  Update the conditional distribution \[p^{(k)}(\hy|y_{r}) = \frac{\exp\left(\delta^{(k)}(\hy,y_{r})\right)  }{\sum_{\hy^{'}}  \exp \left( \delta^{(k)}(\hy^{'},y_{r}) \right) }.\]
\item Update $p^{(k)}(x_1|\hy, x_2)$, $p^{(k)}(x_2|\hy, x_1)$, $p^{(k)}(\hy|x_1)$ and $p^{(k)}(\hy|x_2)$ according to $p^{(k)}(\hy|y_r)$.
\item Update the value of the Lagrangian $\mc L^{(k)}$ with the current distributions.
\item Stop if $\frac{\mc L^{(k)} - \mc L^{(k-1)}}{\mc L^{(k)}} < \epsilon$. Otherwise go to step 4.
\end{enumerate}

\subsection{Convergence of the Algorithm}
\begin{figure}[t]
 \centering
 %\footnotesize
 %\input{./gfx/convergence.tikz.tex}
 %\vspace*{-3mm}
 \includegraphics[width=0.5\textwidth]{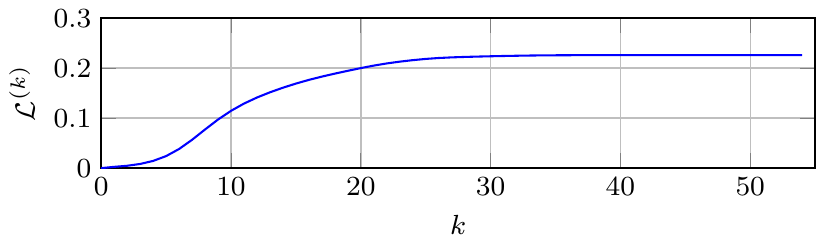}
 \caption{$\mc L^{(k)}$ vs. iteration number $k$.}
 \label{fig:convergence}
 %\vspace*{-3mm}
\end{figure}

Consider the functional 
\begin{align}
&\mc F\left( p(\hy|y_r),t_1(x_1|\hy,x_2), t_2(x_2|\hy,x_1), t_3(\hy|x_1), t_4(\hy|x_2) \right) = \nonumber \\
&-\sum_{x_1}\sum_{x_2} \sum_{\hy} \sum_{y_r} p(\hy|y_r) p(y_r|x_1,x_2) p(x_1)p(x_2) \cdot \left[ \log t_1(x_1|\hy,x_2) + \log t_2(x_2|\hy,x_1) \right]  \nonumber \\
& +\lambda_1 \sum_{x_1} \sum_{y_r}\sum_{\hy}  p(\hy|y_r) p(y_r|x_1) p(x_1) \log\frac{p(\hy|y_r)}{t_3(\hy|x_1)} + \lambda_2 \sum_{x_2} \sum_{y_r}\sum_{\hy}  p(\hy|y_r) p(y_r|x_2) p(x_2) \log\frac{p(\hy|y_r)}{t_4(\hy|x_2)} \label{eq:functional}
\end{align}
for some distributions $t_1(x_1|\hy,x_2)$ on $\mc X_1$, $t_2(x_2|\hy,x_1)$ on $\mc X_2$, $t_3(\hy|x_1)$ on $\mc{\hat{Y}}_r$ and $t_4(\hy|x_2)$ on $\mc{\hat{Y}}_r$. It is clear that if 
\begin{align}
 &t_1(x_1|\hy,x_2) = p(x_1|\hy,x_2)  \nonumber \\
 &t_2(x_2|\hy,x_1) = p(x_2|\hy,x_1)  \nonumber \\
 &t_3(\hy|x_1) = p(\hy|x_1) \nonumber \\
 &t_4(\hy|x_2) =  p(\hy|x_2),\\
 \text{then}\qquad \mc F &= H(X_1|\hY, X_2) + H(X_2|\hY,X_1) \nonumber + \lambda_1 I(Y_r;\hY|X_1) + \lambda_2 I(Y_r;\hY|X_2) \geq 0
\end{align}
and the Lagrangian is 
\[\mc L = H(X_1) + H(X_2) - \mc F.\]
A straightforward adaption of \cite[Lemma 10.8.1]{cover2006elements} gives
\begin{align}
&\arg \min_{t_3(\hy|x_1)} \sum_{x_1} \sum_{y_r}\sum_{\hy}  p(\hy|y_r) p(y_r|x_1) p(x_1) \log\frac{p(\hy|y_r)}{t_3(\hy|x_1)}  = p(\hy|x_1) = \sum_{y_r} p(\hy|y_r) p(y_r|x_1) , \\
 &\arg \max_{t_1(x_1|\hy,x_2)} \sum_{x_1}\sum_{x_2} \sum_{\hy}  p(\hy,x_1,x_2)  \log t_1(x_1|\hy,x_2) 
 = p(x_1|\hy, x_2) = \frac{p(x_1) \sum_{y_r} p(\hy|y_r) p(y_r|x_1,x_2) }{\sum_{x_1^{'}} \sum_{y_r}p(\hy|y_r) p(y_r|x_1^{'},x_2) p(x_1^{'}) }.
\end{align}
The proof is due to the nonnegativity of the information divergence.
It follows that
 \begin{alignat}{2}
 p(\hy|x_2) &=\arg \min_{t_4(\hy|x_2)}  &\mc F(Q,t_1, t_2, t_3, t_4)  \nonumber \\
 p(\hy|x_1) &=\arg \min_{t_3(\hy|x_1)}  &\mc F(Q,t_1, t_2, t_3, t_4)   \nonumber\\
 p(x_2|\hy,x_1) &=\arg \min_{t_2(x_2|\hy,x_1)} &\mc F(Q,t_1, t_2, t_3, t_4)   \nonumber \\
 p(x_1|\hy,x_2) &=\arg \min_{t_1(x_1|\hy,x_2)} &\mc F(Q,t_1, t_2, t_3, t_4) .
\end{alignat}
As a result, one can write the problem
 \begin{align}
 \max_{Q} \mc L(Q,\lambda_1,\lambda_2) &= H(X_1) + H(X_2) 
 - \min_{Q} \min_{t_1} \min_{t_2} \min_{t_3} \min_{t_4} \mc F(Q,t_1, t_2, t_3, t_4) \nonumber
 \end{align}
 as multiple independent minimization problems over the distributions $t_4$, $t_3$, $t_2$, $t_1$ and $Q$. Each minimization with respect to one particular distribution considers the remaining distributions to be fixed.
 The same principle is used for the Blahut-Arimoto algorithm \cite[Chapter 10.8]{cover2006elements} or in \cite{ibm}.
The updates in step 6 and 7 of the iterative algorithm above can be seen as minimization with respect to one distribution given all the other distributions $\mc F$ depends on. It follows that $\mc F$ does not increase at each step and hence $\mc L$ does not decrease.
As $\mc L$ is bounded from above, the sequence $\mc L^{(k)}$ converges to a limit point $\mc L^{(\infty)}$. Under mild conditions, this also implies convergence of $Q$.
The optimizing $Q$ is not unique, as the problem is permutation-symmetric. That is, permuting rows of $Q$ does not affect the value of $\mc L$. The limit point $\mc L^{(\infty)}$ depends on the initial choice of $Q$, but we find that the results do not differ a lot.
Fig.~\ref{fig:convergence} shows that relatively few iterations suffice for convergence.

\subsection{Illustration and Discussion}
\begin{figure}
\centering
 \includegraphics[width=0.45\textwidth]{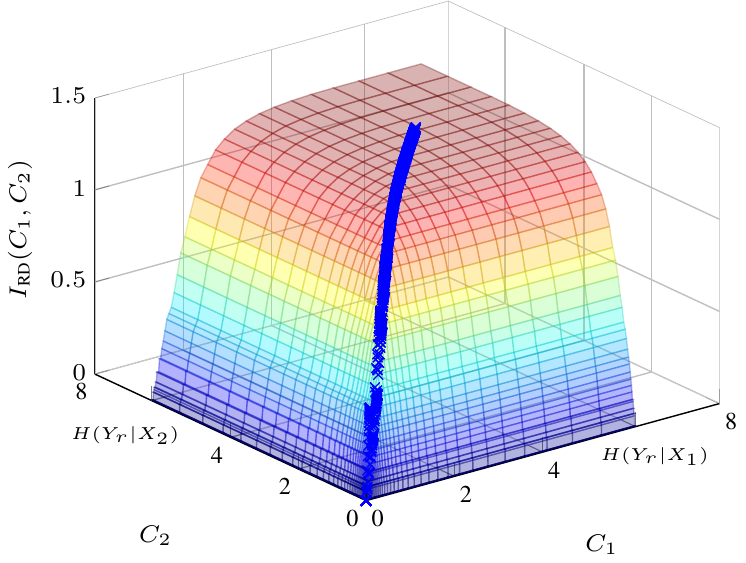}
\caption{$I_{\text{RD}}(C_1,C_2)$ for $0 \leq C_1 \leq H(Y_r|X_1)$, $0 \leq C_2 \leq H(Y_r|X_2)$ and $P_1=1.5$dB, $P_2=4.5$dB, $N_r=0$dB.}
\label{fig:sim}
\end{figure}

We run the algorithm given in the previous section for a system with BPSK modulation at the transmitters, uplink SNRs of $P_1/N_r = 1.5$dB and $P_2/N_r = 4.5$dB and $L= 32$. Fig.~\ref{fig:sim} shows the resulting  $I_{\text{RD}}(C_1,C_2)$-surface. The blue points correspond to the cases where the optimal $Q$ satisfies both inequality constraints with equality. In this case, $Q$ lies on the boundary of both $\mc Q_{C_1}^{(1)}$ and $\mc Q_{C_2}^{(2)}$.
For other points on this surface, one can decrease $C_1$ or $C_2$ without reducing the value of $I_{\text{RD}}$. This is because if $Q$ has to satisfy $I(Y_r;\hY|X_1)\leq C_1$, then the flexibility to obtain particular values for $I(Y_r;\hY|X_2)$ is restricted. We observe that the interval 
\begin{align}
 \left[ \inf_{Q\in\partial {Q}^{(1)}_{C_1}} I(Y_r;\hY|X_2), \sup_{Q\in\partial {Q}^{(1)}_{C_1}} I(Y_r;\hY|X_2)   \right]
\end{align}
with $\partial {Q}^{(1)}_{C_1}=\{ Q \in \Delta_L^{|\mc Y_r|}|I(Y_r;\hY|X_1)= C_1\}$ is very small.
This observation suggests to prefer layered quantization for asymmetric downlink channels, as proposed in \cite{4797715} for Compress \& Forward or in \cite{lim2010layered,do2012layered} for Noisy Network Coding.
Otherwise, the worse user is limiting the quantization accuracy for the better user.

\begin{figure}[Ht]
\centering 
%\footnotesize
%\input{gfx/scalar_q_tikz}
 \includegraphics[width=0.5\textwidth]{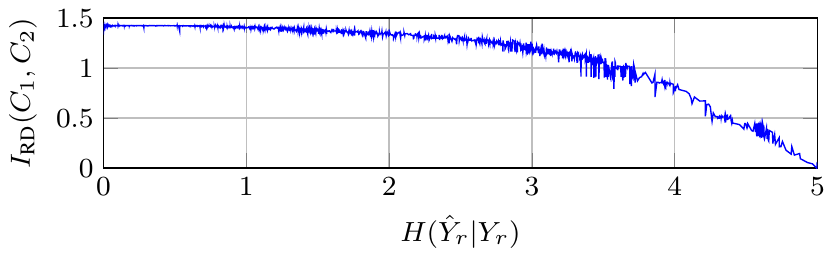}
\vspace*{-3mm}
\caption{$H(\hat{Y}_{r}|Y_{r})$ vs. $I_{\text{RD}}(C_1,C_2)$}
\label{fig:scalar_q}
\end{figure}

Another aspect is the design of quantizers based on the optimal distributions obtained with the proposed algorithm. In general, vector quantizers are needed. From a practical point of view scalar quantizers (where the output of the quantizer depends only on the received symbol, not the whole block of symbols) are interesting. Note that scalar quantizers imply $H(\hY|Y_r)=0$. $H(\hY|Y_r)$  can be used to measure if the conditional pmf $Q$ is close to a scalar quantizer. 
Fig.~\ref{fig:scalar_q} depicts pairs of values for $H(\hY|Y_r)$ and $I_{\text{RD}}(C_1,C_2)$. According to this figure, the distributions with the highest values for $I_{\text{RD}}$ have $H(\hY|Y_r)\approx0$, i.e. the output of the quantizer depends only on the current symbol, not on the whole block. This means that especially in the saturation region of $I_{\text{RD}}$, where $C_1$ and $C_2$ are relatively large, scalar quantizers suffice. Similar to \cite{6620441}, this can be formally shown for $C_1 \geq H(Y_r|X_1)$ and $C_1 \geq H(Y_r|X_2)$. Our numerical results suggest that scalar quantizers suffice already for much smaller values of $(C_1,C_2)$.

\section{Conclusion}
We studied the rate-distortion properties associated with the Quantize-and-Forward scheme of \cite{schnurr2007achievable}.
The function for corresponding rate-distortion tradeoff was be shown to be concave.
We proposed an algorithm to obtain distributions that maximize the sum rate. Although in general the optimal probability distributions imply vector quantizers, scalar quantizers suffice in certain cases.

\section*{Acknowledgments}
The authors are supported by the German Ministry of Education and Research in the framework of the Alexander von Humboldt-Professorship and by the grant DLR@Uni of the Helmholtz Allianz.
The authors thank Gerhard Kramer for his helpful comments.

%\IEEEtriggeratref{2}

\bibliographystyle{IEEEtran}
\bibliography{bib}

\begin{figure}[ht]
\centering
\includegraphics[width=0.3\textwidth]{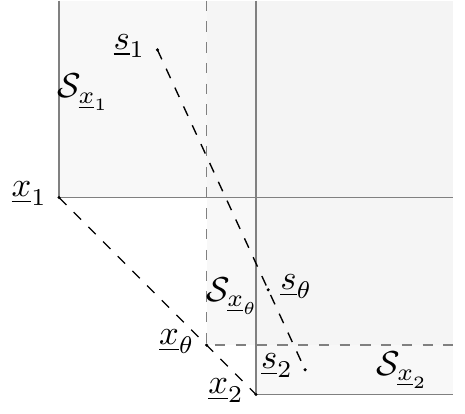}
\caption{Illustration of geometry for Lemma \ref{lemma:fconvexity} for the $2$-dimensional case.}
\label{fig:lemma_illustration}
\end{figure}

\clearpage
\appendix
To prove the concavity property of $I_{\text{RD}}(C_1,C_2)$, we need the following Lemma:
\begin{lemma}
 For a convex function $f(\bar{\ve x})$  ($f: \mathbb{R}^n \rightarrow \mathbb{R}$), the function 
$F(\ve x) = \inf_{\bar{\ve x} \geq \ve x}  f(\bar{\ve x}) $ is convex, where the relation $\bar{\ve x} \geq \ve x$ is component-wise.
\label{lemma:fconvexity}
\end{lemma}
\begin{IEEEproof}
Define the box $\mc S_{\ve x_0} = \{\bar{\ve x}| \bar{\ve x} \geq \ve x_0\}$ and let $\ve x_0^*  = \arg \min _{\bar{\ve x} \in \mc S_{\ve x_0}}  f(\bar{\ve x})$, such that $F(\ve x_0) = f(\ve x_0^*)$.
Choose two points $\ve x_1$, $\ve x_2 \in \mathbb{R}^n$ and let $\ve x_\theta = \theta \ve x_1 + (1-\theta) \ve x_2$, $0\leq \theta \leq 1$.
For any two points $\ve s_1 \in \mc S_{\ve x_1}$, $\ve s_2 \in \mc S_{\ve x_2}$, the point $\ve s_\theta = \theta \ve s_1 + (1-\theta) \ve s_2 $ is an element of $\mc S_{\ve x_\theta}$, with the same choice of $\theta$ as for $\ve x_\theta$. This is true because $\ve s_1 \geq \ve x_1$ and $\ve s_2 \geq \ve x_2$ by definition and thus $\ve s_\theta \geq \theta \ve x_1 + (1-\theta) \ve x_2 = \ve x_\theta$. Therefore, $\ve s_\theta \in \mc S_{\ve x_\theta}$. This is illustrated in Fig.~\ref{fig:lemma_illustration}.
Consequently, for any two optimizers $\ve x_1^* \in \mc S_{\ve x_1}$, $\ve x_2^* \in \mc S_{\ve x_2}$, the point $\ve z = (\theta \ve x_1^* + (1-\theta) \ve x_2^*)$ is in $\mc S_{\ve x_\theta}$. Due to the convexity of $f$, we have 
$f(\ve z) \leq \theta f(\ve x_1^*) + (1-\theta) f(\ve x_2^*)$, and as $\ve z \in \mc S_{\ve x_\theta}$, we have  $f(\ve x_\theta^*) \leq f(\ve z)$.
It follows that $f(\ve x_\theta^*) \leq \theta f(\ve x_1^*) + (1-\theta) f(\ve x_2^*)$ and thus $F(\ve x_\theta) \leq \theta F(\ve x_1) + (1-\theta) F(\ve x_2)$. 
This is true for any $\theta \in [0,1]$, which proves the lemma.
\end{IEEEproof}

\begin{figure}[htbp]
 \centering
\includegraphics[width=0.6\textwidth]{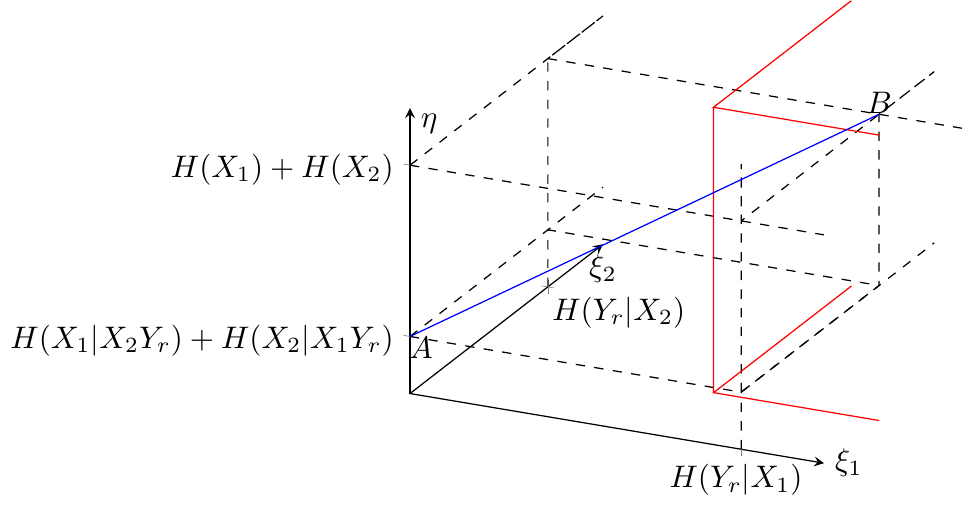}
\caption{Visualization of the box $[0,H(Y_r|X_1)] \times [0,H(Y_r|X_2) \times [H(X_1|X_2 Y_r)+H(X_2|X_1 Y_r), H(X_1) + H(X_2) ] $ containing $\mc C_{\ve{p}_{y_r}}$. The red lines indicate the intersection of the sets $\mc C_{x_1}$ and $\mc C_{x_2}$. }
\label{fig:boxes_convexity_proof}
\end{figure}

The proof for concavity of $I_{\text{RD}}(C_1,C_2)$ is along the lines of \cite{witsenhausen1975conditional} and \cite{zeitler2012} with adaptions to this setup. Recall that we abbreviate $p(\hy|y_r)$ by $Q$. By using the definition of mutual information and dropping the constant terms an equivalent problem (in the sense of the same optimal argument) can be written as 
\begin{align}
 F(x_1,x_2) &:= \inf_{Q}  \left( H(X_1|\hY,X_2) + H(X_2|\hY,X_1)\right), \label{eq:Def_F(x1x2)} \\
\text{s.t. } &H(Y_r|\hY,X_1) \geq x_1, \quad\text{for } 0\leq x_1 \leq H(Y_r|X_1), \nonumber \\
&H(Y_r|\hY,X_2) \geq x_2, \quad\text{for } 0\leq x_2 \leq H(Y_r|X_2).\nonumber
\end{align}
We investigate properties of $F(x_1,x_2)$.
In the following, we often use a vector representation of marginal probability distributions. 
The distribution of a general random variable $Z$ that has cardinality $|\mc Z|$, $p(z)$ is equivalently represented by the column vector $\ve p_{z} \in \Delta_{|\mc{Z}|}$ in the $|\mc{Z}|$-dimensional probability simplex $\Delta_{|\mc{Z}|}$, describing an $(|\mc{Z}|-1)$-dimensional space. 
The $i$-th coordinate is denoted by $p_{z,i} = p(Z=z_i)$. 
Therefore, let $\ve{p}_{y_r} \in \Delta_{|\mc{Y}_r|}$, $\ve{p}_{\hat{y}_r} \in \Delta_{{\yhcard}}$ represent the marginal distribution $p(y_r)$,  $p(\hy)$, respectively. 
Let $B = [\ve b_1, \ldots , \ve b_{{\yhcard}}]$ be a $|\mc{Y}_r| \times {\yhcard}$ stochastic matrix with $\ve b_{i} \in \Delta_{|\mc{Y}_r|}$ in the $i$-th column.
Introduce the random variables $Y_r'$, $X_1'$ and $X_2'$ with respective marginal distributions 
\begin{align}
 \ve p_{y_r'}~&=~\sum_{i=1}^{{\yhcard}}~p(\hat y_{ri}) \ve b_i,&\nonumber \\
 p(x_{1j}')&=~\sum_{i=1}^{|\mc Y_r|} p(x_{1j}|y_{ri}) p(y_{ri}'),&\forall~j=1,\ldots,|\mc X_1| \nonumber \\
 p(x_{2j}') &=   \sum_{i=1}^{|\mc Y_r|} p(x_{2j}|y_{ri}) p(y_{ri}'),&\forall~j=1,\ldots,|\mc X_2|. \nonumber
\end{align}
In general, the matrix $B$ corresponds to $p(y_r'|\hy)$.
Clearly, if $B$ is equal to to $p(y_r|\hy)$, then $p(y_r') = p(y_r)$, $p(x_1') = p(x_1)$ and $p(x_2') = p(x_2)$.
One can write
\begin{align}
 \ve p_{y_r'} &= \sum_{i=1}^{{\yhcard}} p(\hat y_{ri}) \ve b_i  \label{eq:p_proof_schnurr}\\
  \eta  & = \sum_{i=1}^{{\yhcard}} p(\hat y_{ri})  \left[ \underbrace{H(X_1'|X_2', \hY = \hat y_{ri})}_{\leq \log(|\mc{X}_1|)}  + \underbrace{H(X_2'|X_1', \hY = \hat y_{ri})}_{\leq \log(|\mc{X}_2|)} \right]  
       := \sum_{i=1}^{{\yhcard}} p(\hat y_{ri}) g(\ve b_i) \leq \log(|\mc{X}_1|\cdot |\mc{X}_2| )  \\
\xi_1 & =  \sum_{i=1}^{{\yhcard}} p(\hat y_{ri}) H(Y_r'|X_1',\hY = \hat y_{ri}) := \sum_{i=1}^{{\yhcard}} p(\hat y_{ri}) h_{1}(\ve b_i)   \label{eq:xi1_proof} \\
 \\&\text{with } h_{1}(\ve b_i) = H(Y_r'|X_1',\hY = \hat y_{ri}) = -\sum_{y_r'} \sum_{x_1'} p(x_1'|y_r') p(y_r'|\hyri) \log\left( \frac{p(x_1'|y_r') p(y_r'|\hyri)}{\sum_{y_r^{''}} p(x_1'|y_r^{''}) p(y_r^{''}|\hyri)  } \right). \nonumber \\ 
  \xi_2 & =  \sum_{i=1}^{{\yhcard}} p(\hat y_{ri}) H(Y_r'|X_2',\hY = \hat y_{ri})  := \sum_{i=1}^{{\yhcard}} p(\hat y_{ri}) h_{2}(\ve b_i)   \label{eq:xi2_proof_schnurr}
   \\&\text{with } h_{2}(\ve b_i) = H(Y_r'|X_2',\hY = \hat y_{ri}) = -\sum_{y_r'} \sum_{x_2'} p(x_2'|y_r') p(y_r'|\hyri) \log\left( \frac{p(x_2'|y_r') p(y_r'|\hyri)}{\sum_{y_r^{''}} p(x_2'|y_r^{''}) p(y_r^{''}|\hyri)  } \right).  \nonumber
\end{align}
The problem in (\ref{eq:Def_F(x1x2)}) can be stated as
\begin{align}
F(x) =  \inf_{\ve p_{y_r'} = \ve{p}_{y_r}} \eta, \qquad \text{s.t. } \xi_1 \geq x_1, \quad \xi_2 \geq x_2.
\end{align}
Define the mapping $\ve b_i \in \Delta_{|\mc{Y}_r|} \rightarrow \left(\ve b_i, h_{1}(\ve b_i), h_{2}(\ve b_i), g(\ve b_i) \right)$.
Remember that $\Delta_{|\mc{Y}_r|}$ is $(|\mc{Y}_r|-1)$-dimensional, so the polytope $\Delta_{|\mc{Y}_r|} \times [0, \log(|\mc{Y}_r|)] \times [0, \log(|\mc{Y}_r|)] \times [0, \log(|\mc{X}_1|\cdot |\mc{X}_2|)]$ is $(|\mc{Y}_r|+2)$-dimensional and the mapping assigns points inside this polytope for each choice of $\ve b_i$.
Let $\mc S$ be the set of all such points for all possible $\ve b_i$. As $h_{1}(\ve b_i)$, $h_{2}(\ve b_i)$ and $g(\ve b_i)$ are continuous functions of $\ve b_i$ \cite[Chapter 2.3]{yeung2008information}, $\mc S$ is compact and connected. 
Define $\mc C$ as the convex hull of $\mc S$, i.e., $\mc C = \text{conv}(\mc S)$.
By definition of the convex hull, the set of pairs $(\ve p_{y_r'}, \xi_1, \xi_2, \eta)$ defined in (\ref{eq:p_proof_schnurr}) - (\ref{eq:xi2_proof_schnurr}) form $\mc C$, for all integers ${\yhcard}>0$, $\ve{p}_{\hat{y}_r} \in \Delta_{{\yhcard}}$, $\ve b_i \in \Delta_{|\mc{Y}_r|}$, $i = 1,\ldots,{\yhcard}$.
By the Fenchel-Eggleston strengthening of Carath\'{e}odory's theorem \cite[Appendix A]{el2011network}, every point in $\mc C$ can be obtained by taking a convex combination of at most ${\yhcard}\leq |\mc{Y}_r|+2$ points from the set $\mc S$.
\begin{prop}
 The function $F(x_1,x_2)$ is jointly convex in $x_1$, $x_2$, for $0\leq x_1 \leq H(Y_r|X_1)$ and $0\leq x_2 \leq H(Y_r|X_2)$.
\label{prop:F(x1x2)convex}
\end{prop}
\begin{IEEEproof}
The function $F(x_1,x_2)$ is the minimum of $\eta$ for which $\ve p_{y_r'} = \ve{p}_{y_r}$ and $\xi_1 \geq x_1$, $\xi_2 \geq x_2$. 
Define $\mc C_{\ve{p}_{y_r}}$ as the the projection of the intersection of $\mc C$ with the convex and compact set defined by $\ve p_{y_r'} = \ve{p}_{y_r}$ onto the 3-dimensional space $(\xi_1,\xi_2,\eta) \subset \mathbb{R}^3$.
That is, $\mc C_{\ve{p}_{y_r}} = \text{Proj}\{\mc C \cap \mc L_{\ve{p}_{y_r}} \}$, where 
\[\mc L_{\ve{p}_{y_r}} = \{(\ve p_{y_r'},\xi_1, \xi_2,\eta) \subset \mathbb{R}^{|\mc{Y}_r|+2}| \ve p_{y_r'} = \ve{p}_{y_r}\}.\]
$\mc L_{\ve{p}_{y_r}}$ is convex and compact.
As convexity is preserved under intersection \cite[Sect. 2.3.1]{boyd2004convex} and projection onto coordinates \cite[Sect. 2.3.2]{boyd2004convex}, the set $\mc C_{\ve{p}_{y_r}}$ is also convex and compact.
Now define the convex and compact sets 
\begin{align}
\mc C_{x_1} =& \left\{ (\xi_1,\xi_2,\eta) \in \mathbb{R}^3 | \xi_1 \geq x_1 \right\}, \nonumber \\ 
\mc C_{x_2} =& \left\{ (\xi_1,\xi_2,\eta) \in \mathbb{R}^3 | \xi_2 \geq x_2 \right\}, \nonumber \\
\text{and let } \mc C_{\ve{p}_{y_r},x_1,x_2} :=& C_{\ve{p}_{y_r}} \cap \mc C_{x_1} \cap \mc C_{x_2},
\end{align}
 which is convex and compact \cite[Sect. 2.3.1]{boyd2004convex}.
That is, the infimum in (\ref{eq:Def_F(x1x2)}) can be attained and is thus a minimum, if 
 $\mc C \cap \mc L_{\ve{p}_{y_r}} \not = \emptyset$ and $C_{\ve{p}_{y_r}} \cap \mc C_{x_1} \cap \mc C_{x_2} \not = \emptyset$.
Both requirements can be shown using the same argument:
$\eta$ takes on its maximal value $H(X_1) + H(X_2)$ if and only if $Y_r$ is independent of $\hY$, so for $Y_r \perp \hY$. We can achieve this by chosing ${\yhcard}=1$, $\ve b_1 = \ve{p}_{y_r}$, $p(\hat y_{r1})=1$ like in \cite{witsenhausen1975conditional}.  It follows that $\ve p_{y_r'} = \ve{p}_{y_r}$, $\xi_1 = H(Y_r|X_1)$ and $\xi_2 = H(Y_r|X_2)$. One obtains the upper right corner of the box with the coordinates $(\xi_1,\xi_2,\eta) = (H(Y_r|X_1), H(Y_r|X_2),H(X_1)+H(X_2))$, labeled with $B$ in Fig. \ref{fig:boxes_convexity_proof}.
$\eta$ takes on its minimal value $H(X_1|X_2 Y_r)+H(X_2|X_1 Y_r)$ if there is a bijective mapping between $Y_r$ and $\hY$. Choosing ${\yhcard}=|\mc{Y}_r|$, $\ve{p}_{\hat{y}_r} = \ve{p}_{y_r}$, $B = [\ve b_1, \ldots, \ve b_{|\mc{Y}_r|}]$ as the identity matrix, it follows that $\ve p_{y_r'} = \ve{p}_{y_r}$, $\xi_1 = \xi_2 =0$.
We obtain the lower left corner of the box with coordinates $( 0,0,H(X_1|X_2 Y_r)+H(X_2|X_1 Y_r))$, labeled with $A$.
By the convexity of $\mc C_{\ve{p}_{y_r}}$, the straight line connecting points $A$ and $B$ (marked in blue in Fig. \ref{fig:boxes_convexity_proof}) must lie inside $\mc C_{\ve{p}_{y_r}}$.
Therefore, the intersection $C_{\ve{p}_{y_r}} \cap \mc C_{x_1} \cap \mc C_{x_2}$ is never empty, as for each pair of $x_1$ and $x_2$ satisfying $0\leq x_1\leq H(Y_r|X_1)$, $0\leq x_2 \leq H(Y_r|X_2)$
 the straight blue line between $A$ and $B$ and the red box defined by $\mc C_{x_1} \cap \mc C_{x_2}$ have points in common.

Define the lower boundary of the set $\mc C_{\ve{p}_{y_r}}$ as $f(x_1,x_2)$. Its domain is the projection of $\mc C_{\ve{p}_{y_r}}$ to the $(\xi_1,\xi_2)$-plane, i.e.:  
\[ \dom f = \left\{ (x_1,x_2) \Big| \inf_\eta \{ (\xi_1,\xi_2,\eta) \in \mc C_{\ve{p}_{y_r}} \big| \xi_1=x_1, \xi_2=x_2 \} < \infty \right\}.\]
As $\mc C_{\ve{p}_{y_r}}$ is convex, $f(x_1,x_2)$ is a convex function inside its domain.
Define $\tilde f(x_1,x_2)$ as the extended-value extension of $f$ \cite[Section 3.1.2]{boyd2004convex}:
\begin{align}
 \tilde f(x_1,x_2) = \begin{cases}
                              f(x_1,x_2) & (x_1,x_2) \in \dom f \\
                              \infty & (x_1,x_2) \not \in \dom f.
                             \end{cases} 
\end{align}
$\tilde f$ is now defined inside the whole box $0\leq x_1\leq H(Y_r|X_1)$, $0\leq x_2 \leq H(Y_r|X_2)$. Convexity of $f$ implies convexity of $\tilde f$ inside this box.
$F(x_1,x_2)$ can be defined as 
\begin{align}
 F(x_1,x_2) = \inf \left\{ \tilde f(\bar x_1,\bar x_2)| \bar x_1\geq x_1, \bar x_2 \geq x_2 \right\}.
 \label{eq:Fandf}
\end{align}
By Lemma \ref{lemma:fconvexity}, convexity of $\tilde f(x_1,x_2)$ implies convexity of $F(x_1,x_2)$.
\end{IEEEproof}
\begin{cor}
 $I_{\text{RD}}(C_1,C_2)$ is a concave function in $(C_1,C_2)$, for $0\leq C_1 \leq  H(Y_r|X_1)$, $0\leq C_2 \leq  H(Y_r|X_2)$.
\label{cor:I(C1C2)_concave}
\end{cor}
The proof follows by Proposition \ref{prop:F(x1x2)convex}.

\end{document}